\documentstyle[aps,pra,epsfig,twocolumn]{revtex}

\def\be{\begin{equation}}
\def\ee{\end{equation}}
\def\bea{\begin{eqnarray}}
\def\eea{\end{eqnarray}}
\def\bma{\begin{mathletters}}
\def\ema{\end{mathletters}}
\def\C{{\cal C}}
\def\N{{\cal N}}

\newcommand{\bra}[1]{\mbox{$\langle #1 |$}}
\newcommand{\ket}[1]{\mbox{$| #1 \rangle$}}
\newcommand{\braket}[2]{\mbox{$\langle #1  | #2 \rangle$}}
\newcommand{\proj}[1]{\ket{#1}\!\bra{#1}}

\begin{document}
\draft

\title{
When only two thirds of the entanglement can be distilled\\
}

\author{G. Vidal and J. I. Cirac}

\address{Institut f\"ur Theoretische Physik, Universit\"at Innsbruck,
A-6020 Innsbruck, Austria}

\date{\today}

\maketitle

\begin{abstract}
We provide an example of distillable bipartite mixed state such that, even in the asymptotic limit, more pure-state entanglement is required to create it than can be distilled from it. Thus, we show that the irreversibility in the processes of formation and distillation of bipartite states, recently proved in [G. Vidal, J.I. Cirac, Phys. Rev. Lett. 86, (2001) 5803-5806], is not limited to bound-entangled states.

\end{abstract}


\narrowtext

Distillation is one of the basic concepts in entanglement theory. As shown in the pioneering works on entanglement transformations \cite{Be961,Be960}, it is possible to use local operations and classical communication (LOCC) to convert, in the asymptotic limit ($N\rightarrow \infty$), $N$ copies of some bipartite mixed state $\rho$ into $M$ of copies of some reference pure state, the maximally entangled state
\be
\ket{\Phi} \equiv \frac{1}{\sqrt{2}}(\ket{00}+\ket{11})
\ee
of a two-qubit system, which is said to contain one ebit (entangled bit). Moreover, the distillable entanglement $E_{D}(\rho)$, defined as the maximal achievable yield $M/N$, was shown to be often finite. This is a remarkable result with important implications in quantum information theory. It says, for instance, that a noisy channel can be used to establish perfect quantum communication between two distant parties, if these are allowed to perform LOCC. Indeed, the imperfect channel can be used to create many copies of some mixed entangled state $\rho$, which can then be purified into fewer copies of $\ket{\Phi}$ and subsequently used to achieve perfect quantum communication through teleportation \cite{Be93}.
 
A notion dual to distillation is that of preparation of $\rho$ using pure-state entanglement and LOCC \cite{Be960}. Now $M$ copies of $\ket{\Phi}$ are transformed into $N$ copies of $\rho$. The entanglement cost $E_{C}(\rho)$ \cite{Ha00} (asymptotic version of the entanglement of formation $E_F(\rho)$ \cite{Be960,Wo98}) is defined as the minimal ratio $M/N$ asymptotically achievable by LOCC. $E_C(\rho)$ quantifies the amount of pure-state entanglement required to create a copy of $\rho$, in the above asymptotic sense.

Notice that the processes of formation and distillation can be concatenated into a cycle. Starting from $NE_{C}(\rho)$ copies of $\ket{\Phi}$, two distant parties can use LOCC to prepare $N$ copies of $\rho$; and the $N$ copies of $\rho$ can be subsequently distilled back into $NE_{D}(\rho)$ copies of $\ket{\Phi}$,
\be
\ket{\Phi}^{\otimes NE_C(\rho)} \Rightarrow \rho^{\otimes N} \Rightarrow \ket{\Phi}^{\otimes NE_D(\rho)}.
\ee
Already in the early contributions it was suggested that maybe sometimes this cycle can not be closed completely, in that perhaps not all the initial pure-state entanglement used in the preparation process can be recovered through distillation. That is, maybe an irreversible loss of quantum correlations takes place during the mixing of pure-state entanglement into $\rho^{\otimes N}$ and, accordingly, the distillable entanglement $E_D(\rho)$ is smaller than the entanglement cost $E_D(\rho)$. 

Very recently this phenomenon has been proved to indeed occur \cite{Vi01}. In particular, it has been shown that some undistillable bipartite state $\rho_b$ ---i.e. with $E_D(\rho_b) = 0$ ebits--- has non-vanishing entanglement cost. Notably, the irreversibility observed in the asymptotic preparation and distillation of $\rho_b$ remains even when LOCC are supplemented with loaned pure-state entanglement, to be returned after the manipulation, in the so-called catalytic LOCC setting.

The results in \cite{Vi01} still leave, however, an important question open.
One could associate the irreversibility demonstrated there to the fact that the state $\rho_b$ is bound entangled, that is, to the remarkable property that no pure-state entanglement at all can be distilled from it \cite{Ho98}. 
It could well be the case that the gap observed between $E_C$ and $E_D$ is just a characteristic feature of some bound entangled states, whereas $E_C = E_D$ always holds for distillable states. After all, this is the case for bipartite pure states \cite{Be96} and some simple cases of mixed state \cite{trivial}, which exhaust all the cases where $E_C$ and $E_D$ have been computed.

In this paper we will present an example of bipartite mixed state $\sigma$ that can be distilled, that is $E_D(\sigma) > 0$, and such that $E_C(\sigma) > E_D(\sigma)$. We extend thereby the irreversibility result of \cite{Vi01} to the case of distillable states. In particular, the extension also holds for catalytic LOCC transformations. 

A widely recognized, major problem concerning the study of mixed-state entanglement is that it is very difficult to compute the asymptotic measures $E_C$ and $E_D$. Here, however, we are not interested in the actual values of $E_C(\sigma)$ and $E_D(\sigma)$. For the present purposes it is sufficient to show that $\sigma$ can be distilled, and to bound $E_D(\sigma)$ and $E_C(\sigma)$ tight enough from above and from below, respectively, so that the bounds already imply a gap between the two quantities. We start by collecting an amalgam of useful facts. 

($i$) A sufficient condition for a mixed state $\rho$ to be distillable is that a projector $P$ into a $\C^2 \otimes \C^2$ subspace (that is, a subspace which is the tensor product of two-dimensional subspaces for each of the two separated parts of the composite system) exists such that the projection $P \rho P^{\dagger}$ is still entangled \cite{Ho98}, that is, such that the partial transposition of $P \rho P^{\dagger}$ has a negative eigenvalue.

($ii$) The logarithmic negativity $E_N(\rho) \equiv \log_2 (1+2\N(\rho))$ \cite{ViWe}, where $\N(\rho)$ is the absolute value of the sum of negative eigenvalues of partial transposition of $\rho$, is an upper bound to the distillable entanglement. Werner's bound reads \cite{ViWe}, 
\be
E_D(\rho) \leq E_N(\rho).
\ee
In addition, $E_N$ is an additive function,
\be
E_N(\rho_1\otimes\rho_2)=E_N(\rho_1) + E_N(\rho_2),
\ee 
which only vanishes for states with positive partial transposition (PPT states),
\be
E_N(\rho_{PPT})=0.
\ee
Finally, and very important to us, $E_N(\rho)$ is a continuous function of $\rho$.

($iii$) The entanglement of formation $E_F$ \cite{Be960} of $\rho$ is bounded below by \cite{Vi01}
\be
E_F(\rho) \geq -\log_2 \alpha,
\ee
where $\alpha$ is the maximal overlap of a product state $\ket{a~b}$ with the projector $\Pi$ onto the support of $\rho$,
\be
\alpha \equiv \max_{\ket{a~b}} \bra{a~b} \Pi \ket{a~b}.
\ee
Accordingly, the entanglement cost $E_C(\rho)$ is bounded below by \cite{Vi01}
\be
E_C(\rho) \geq -´log_2 \beta,
\ee
if for all $N$ the maximal overlap of a normalized product vector $\ket{a_N~b_N}$ with the $N$-fold tensor product of $\Pi$ is at most $\beta^N$,
\be
\max_{\ket{a_N~b_N}}\bra{a_N~b_N} \Pi^{\otimes N} \ket{a_N~b_N} \leq \beta^N.
\label{maxoverN}
\ee

($iv$) The four-dimensional subspace $V \subset \C^3\otimes\C^3$ orthogonal to
the five product vectors
\bea
|0\rangle &\otimes& (|0\rangle +|1\rangle),\nonumber\\
(|0\rangle+|1\rangle) &\otimes& |2\rangle,\nonumber\\ |2\rangle
&\otimes& (|1\rangle +|2\rangle),\nonumber\\
(|1\rangle+|2\rangle)&\otimes& |0\rangle,\nonumber\\
(|0\rangle-|1\rangle+|2\rangle) &\otimes& (|0\rangle
-|1\rangle+|2\rangle), \label{prod}
\eea
does not contain product vectors \cite{Be99}. The projector $\Pi_b$ onto $V$ satisfies: ($a$) it has a PPT \cite{Be99}, ($b$) it fulfills Eq. (\ref{maxoverN}) with $\beta < 0.99$ \cite{Vi01}.

We introduce now a one-parameter family of states
\be
\sigma(p) \equiv (1-p) \rho_b + p \proj{\psi},
\label{sigma}
\ee
where $\rho_b \equiv \Pi_b/4$ is the PPT bound entangled state introduced in \cite{Be99} in the context of the so-called non-extendible product basis, and used in \cite{Vi01} to prove irreversibility of asymptotic manipulations, and 
\be
\ket{\psi}\equiv \frac{1}{\sqrt{6}}(\ket{00}-\ket{01}-2\ket{11})
\label{estatpur}
\ee
is an entangled pure state that is orthogonal to all product states of Eq. (\ref{prod}), that is $\ket{\psi} \in V$. For $p=0$ we recover $\rho_b$, for which we know that $E_C(\rho_b) > -\log_2 0.99 > D(\rho_b)=0$. In what follows we will use facts ($i$)-($iv$) and perturbation theory to show that for $p>0$ we encounter states $\sigma(p)$ which can be distilled, and with $E_C(\sigma_p) > E_D(\sigma_p)$.

The family of states $\sigma(p)$ in Eq. (\ref{sigma}) has been carefully chosen to fulfill two important properties.
First, $\sigma(p)$ is supported on $V$, since $V$ is the support of $\rho_b$ and also $\ket{\psi}$ is supported in $V$. Using ($iii$) and ($iv.b$) this means that for any $p \in [0,1]$ we have a constant lower bound for $E_C$.

{\bf Property 1:} The entanglement cost of $\sigma(p)$, $p\in [0,1]$, is bounded below by
\be
E_C(\sigma(p)) > -log_2 ~0.99 =0.015 ~\mbox{ebits}.
\label{lowerbound}
\ee

Let $\rho_b^{T_A}$ denote the partial transposition of $\rho_b$, and $P$ the rank-four, product projector $(\proj{0}+\proj{1})\otimes(\proj{0}+\proj{1})$. Notice that, by construction, $\rho_b^{T_A}=\rho_b\geq 0$, $(P\rho_bP^{\dagger})^{T_A} = P\rho_b P^{\dagger}$, and $P\ket{\psi}=\ket{\psi}$. The second important feature of $\sigma(p)$ is that, for any $p>0$ the partial transposition of the projection $P\sigma(p) P^{\dagger}$, 
\be
(P\sigma(p)P^{\dagger})^{T_A} = (1-p)P\rho_bP^{\dagger} + p\proj{\psi}^{T_A}, 
\label{trans}
\ee
has a negative eigenvalue $n$. Therefore, because of fact ($i$), the corresponding state $\sigma(p)$ can be distilled. 

{\bf Property 2:} For $p\in (0,1]$, the state $\sigma(\rho)$ can be distilled, that is
\be
E_D(\sigma(p)) > 0 \mbox{ ebits.}
\ee 
For instance, for $p=0.015$, $|n| = 2.7\times 10^{-4}$ (see also Fig. (1)).

 Property 2 has been achieved by selecting a projector $P$ such that $(P\rho_bP^{\dagger})^{T_A}$ has only rank three, and thus one vanishing eigenvalue, whereas $\ket{\psi}\in V$ has been chosen so that $(P\proj{\psi}P^{\dagger})^{T_A} = \proj{\psi}^{T_A}$, that is, so that the negative eigenvalue of $\proj{\psi}^{T_A}$ entirely contributes to (\ref{trans}). We can use perturbation theory to check what is the effect of such choices.

Let $M$ and $N$ be finite dimensional hermitian operators, $M=\sum_{i=0}^l m_i\proj{m_i}$ the spectral decomposition of $M$, with $m_i$ its decreasingly ordered eigenvalues and $m_0\neq m_1$, and let $\epsilon$ be a small parameter. Then the lowest eigenvalue of $M+\epsilon N$ is, as given by perturbation theory \cite{saku},
\be
m_0 + \epsilon\bra{0}N\ket{0} + \epsilon^2 \sum_{i=1}^l \frac{|\bra{m_0}M\ket{m_0}|^2}{m_0-m_i} + {\cal O}(\epsilon^3).
\ee
Making the proper identifications we realize that the negative eigenvalue $n(p)$ of the operator in Eq. (\ref{trans}) is
\be
n = -|k|p^2 + {\cal O}(p^3),
\ee
where $|k|>0$ is of the order of $1$ and the zero and first order contributions vanish due, respectively, to the fact that the smallest eigenvalue of Eq. (\ref{trans}) vanishes, and to the fact that the corresponding eigenvector, $\ket{\tau} \equiv \ket{0}\otimes(\ket{0}+\ket{1})/\sqrt{2}$, fulfills
\be
\bra{\tau} (\proj{\psi}^{T_A}) \ket{\tau} = \braket{\tau}{\psi} \braket{\psi}{\tau}= 0.
\ee
Finally, for $p\leq 1$ such that contributions ${\cal O}(p^3)$ may become important, numerical calculations show that $|n|$ grows monotonically with $p$ (see Fig. (1)).
  
Summarizing, so far we have learn that $\sigma(p)$ can be distilled for any $p>0$, while the entanglement cost is bounded below by Eq. (\ref{lowerbound}). In order to complete the result we need to prove that the distillable entanglement of $\sigma(p)$ is, in some regime of $p\in (0,1]$, smaller than the lower bound (\ref{lowerbound}). This would already follow from the above if $E_D(\sigma(p))$ were a continuous function of $p$. For $p=0$ we have the bound entangled state $\rho_b$, that is, $E_D(\sigma(0))=0$ ebits, whereas at the other extreme, $p=1$, we have the pure entangled state $\ket{\psi}$, whose distillable entanglement $E_D$ (and entanglement cost $E_C$) can be easily computed and reads $E_D(\sigma(1))=0.55$ ebits. But, unfortunately, we can not base our argument in the continuity of $E_D(\sigma(p))$ as a function of $p$, to conclude that an intermediate $p$ must exist such that the distillable entanglement is non-zero and still below the bound (\ref{lowerbound}). Whereas it may well be that $E_D(\rho)$ is a continuous function of $\rho$, this has not been proved. Notice that a plausible objection to continuity relies on the fact that $E_D(\rho)$ is actually a function of $\rho^{\otimes N}$ in the large $N$ limit. Therefore, a small perturbation of $\rho$, which produces a large deviation in $\rho^{\otimes N}$, may imply a discontinuous change in $E_D(\rho)$.

Nevertheless, following fact ($ii$), the logarithmic negativity $E_N(\sigma(p))$ is a continuous upper bound for $E_D(\sigma(p))$ (see Fig. (1)). A direct calculation of $E_N(\sigma(p))$ finally proves the irreversibility of the preparation-distillation cycle for distillable bipartite states. In particular, for $p=0.0015$ we have
\be
E_D(\sigma(0.0015)) < E_N(\sigma(0.0015)) = 0.012 \mbox{ ebits}.
\ee
Thus, $\sigma(0.0015)$ is an example of distillable state with a finite gap $E_C - E_D > 0.003$ ebits. 

We can now further use the properties of the logarithmic negativity $E_N$ to show that such a gap remains even when pure-state entanglement is loaned to assist in the transformations, as it was done with $\rho_b$ in \cite{Vi01}. This is achieved by considering a distillation process starting from $N$ copies of $\sigma(p)$ together with $L$ copies of $\ket{\Phi}$,
\be
\sigma^{\otimes N}\otimes\proj{\Phi}^{\otimes L} \Rightarrow \ket{\Phi}^{\otimes L+NE_D^c(\sigma)},
\ee
Where $E_D^c(\sigma)$ denotes the distillable entanglement of $\sigma$ in the catalytic setting.
 For any $N$ and $L$, we have
\bea
E_N(\sigma^{\otimes N}\otimes \proj{\Phi}^{\otimes L})&=& E_N(\sigma^{\otimes N})+ E_N(\Phi^{\otimes L})\\ &=& NE_N(\sigma)+L,
\eea
where we have used additivity of $E_N$ and the fact that $E_N(\Phi)=1$. This means that even in the large $N$ limit, and once the $L$ loaned states $\ket{\Phi}$ have been discounted from the distillation outcome, at most $NE_N(\sigma)$ ebits of entanglement has been distilled, so that even in the catalytic scenario the bound $E^c_D(\sigma) < E_N(\sigma)$ holds. Therefore the irreversibility result of the paper also applies to this case.

We have shown that the irreversibility in the asymptotic manipulation of bipartite mixed states is not a phenomena restricted to bound entangled states, by providing an specific example of distillable state with a finite gap between its entanglement cost $E_C$ and its distillable entanglement $E_D$. Notice that these results legitimate the use of different measures of entanglement, such as $E_C$ and $E_D$, to quantify, in the asymptotic limit, the resources of entangled mixed-states. The search for an intrinsic irreversibility in the asymptotic manipulation of bipartite systems has motivated, through several contributions ---see for instance \cite{Vi01,ViWe,Martin,Conjec,Dur,Divi}---, the development of many techniques for the study of entanglement and has certainly implied an important gain in insight. Paradoxically, a remaining open question is now whether a non-trivial example \cite{notrivial} of bipartite mixed states exist for which the processes of preparation and distillation can be performed in a fully reversible fashion.

The authors acknowledge motivating discussions with C.H. Bennett, D.P. Divincenzo, G. Giedke, D.W. Leung, J. Preskill, J.A. Smolin, B.M. Terhal and R.F. Werner. G.V. thanks C.H. Bennett for the invitation to visit his group at IBM, May 2001, and all the members of the group for hospitality during the visit. This work was supported by the Austrian Science Foundation under the SFB ``control and measurement of coherent quantum systems'' (Project 11), the European Community under the TMR network ERB--FMRX--CT96--0087, project EQUIP (contract IST-1999-11053), and contract HPMF-CT-1999-00200, the European Science Foundation, and the Institute for Quantum Information GmbH.

\begin{figure} 
\epsfysize=6.5cm
\begin{center}
 \epsffile{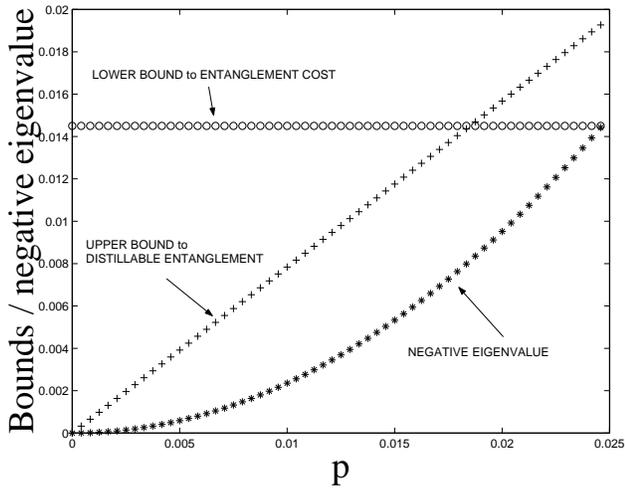}
\end{center}
 \caption{Finite gap between entanglement cost $E_C$ and distillable entanglement $E_D$ for distillable states. We obtain, as a function of $p$, an upper bound $E_N(\sigma(p))$ (diagonal line) for $E_D(\sigma(p))$ which in some regime is smaller than the lower bound (horizontal line) for the entanglement cost $E_C(\sigma(p))$. Both bounds are expressed in ebits. The lower curve corresponds to $20|n|$, where $|n|$ is the modulus of the negative eigenvalue of the operator $(P\sigma(p) P^{\dagger})^{T_A}$, and indicates that the distillable entanglement $E_D(\sigma(p))$ is finite.}
\label{fig:figura}\nonumber 
\end{figure}


\begin{references}
\bibitem{trivial} The entanglement cost $E_C$ and the distillable entanglement $E_D$ can also be computed and shown to coincide in some simple cases. Consider, for instance, the bipartite state
\be
\rho=\sum_i p_i \proj{\Phi_i},
\ee
where each $\ket{\Phi_i} \equiv ( \ket{\phi_i~\psi_i} + \ket{\phi_i^{\perp}\psi_i^{\perp}} )/\sqrt{2}$ satisfies that $tr(\rho_{A_i}\rho_{A_j}) = 0$ when $i\neq j$, for $\rho_{Ai}\equiv$ tr$_B\proj{\Phi_i}$. Then from one copy of $\rho$ we can distill one ebit of entanglement by using a local measurement in part $A$ of the system, given by projectors $E_i$ onto the range of $\rho_{A_i}$, since $E_i\otimes I_B \rho E_i^{\dagger}\otimes I_B = p_i\proj{\Phi_i}$. At the same time, one ebit of entanglement is manifestly sufficient to create a copy of $\rho$. Thus, asymptotically we have $E_{C}(\rho) \leq 1$ ebit and $E_D(\rho) \geq 1$ ebit, but since entanglement cannot be created using LOCC, we have $E_C\geq E_D$, which implies $E_D(\rho)=E_C(\rho)=1$ ebit.
\bibitem{notrivial} By non-trivial example we mean a bipartite mixed states whose density matrix has full rank.
\end{references}
\end{document}